\documentclass[10pt,twoside]{article}
\usepackage{color}
\usepackage{amsmath}
\usepackage{amsfonts}
\usepackage{amssymb}
\usepackage{mathtools}
\usepackage{todonotes}
\DeclareMathOperator{\sgn}{sgn}

\usepackage{8OSME-style}

\title{Nonlinear Kinematics of Recursive Origami Inspired by the Spidron}
\authorlist[Imada, Hull, Ku, Tachi]{Rinki Imada, Thomas C. Hull, Jason S. Ku, Tomohiro Tachi} 
\affiliations{Rinki Imada \at The University of Tokyo, Japan, \email{r-imada@g.ecc.u-tokyo.ac.jp}  \and Thomas C. Hull \at Franklin \& Marshall College, United States, \email{thull1@fandm.edu} \and Jason S. Ku \at National University of Singapore, Singapore, \email{jasonku@nus.edu.sg} \and Tomohiro Tachi \at The University of Tokyo, Japan, \email{tachi@idea.c.u-tokyo.ac.jp}
}

\makeatletter
  \let\runtitle\@title
  \let\runauthor\shortauthor
\makeatother

\begin{document}

\maketitle

\begin{abstract}
    Non-periodic folding of periodic crease patterns paves the way to novel nonlinear phenomena that cannot be feasible through periodic folding. 
    This paper focuses on the non-periodic folding of recursive crease patterns generalized from Spidron.
    Although it is known that Spidron has a $1$-DOF isotropic rigid folding motion, its general kinematics and dependence on the crease pattern remain unclear.
    Using the kinematics of a single unit cell of the Spidron and the recursive construction of the folded state of multiple unit cells, we consider the folding of the Spidron that is not necessarily isotropic.
    We found that as the number of unit cells increases, the non-periodic folding is restricted and the isotropic folding becomes dominant.
    Then, we analyze the three kinds of isotropic folding modes by constructing $1$-dimensional dynamical systems governing each of them.
    We show that these systems can possess different recursive natures depending on folding modes even in an identical crease pattern.
    Furthermore, we show their novel nonlinear nature, including the period-doubling cascade leading to the emergence of chaos.
\end{abstract}

\section{Introduction}\label{sec:introduction}
The kinematics of rigid origami whose crease patterns have certain symmetries has promoted the development of origami science and engineering.
Among such crease patterns, origami that have translation symmetries, such as Miura-ori, have attracted much attention.
In particular, \emph{uniform folding} of periodic crease patterns, where the folding of each unit cell is identical, has been well-studied~\cite{schenk2013geometry,tachi2015rigid,feng2020helical,mcinerney2020hidden}.
This uniformity in folding makes their kinematics easier to understand and control; however, it neglects possible nonlinearity that may be present when they are folded without uniformity. 

Recently, the non-uniform folding of periodic origami crease patterns has gained attention as a source of novel nonlinear phenomena, including the localization of deformations~\cite{chen2016topological} and the undulation in tubular origami tessellations~\cite{imada2022geometry,imada2023undulations,reddy2023frustration}.
Because of the high complexity of non-uniform folding, researchers have used various dynamics- or kinematics-based simulations for their analysis.
These simulations help us know how non-uniform folded structures behave, but not \textit{why} they behave this way.
Hence, the broad range of mathematics behind the non-uniform folding remains unclear.
To approach the system behind the behavior of non-uniform folding, the authors have applied a \emph{dynamical systems} model to periodic origami patterns~\cite{imada2022geometry,imada2023undulations}.
The dynamical system refers to a recurrence relation based on the coupled folding motion between adjacent unit cells, which can capture the nonlinear nature of non-uniform folding.
\cite{imada2023undulations} points out that the existence of scaling of modules in the pattern will drastically change the system's property; specifically, the system loses conservativeness when the pattern is generated by scaling transformation.
However, the actual behavior of patterns with scaling transformation was not fully investigated.
This paper studies generalized Spidrons as examples of origami patterns with scale transformations.

The \emph{Spidron} is a self-similar triangular pattern originally proposed by Dániel Erdély~\cite{erdely2004spidron}.
It is known that the Spidron has a $1$ degree of freedom (DOF) rigid folding motion~\cite{szilassi2004right,kiss2018spidron}.
Significantly, while the Spidron pattern has self-similarity, its rigid folded state does not necessarily have a self-similarity.
This \emph{non-self-similar folding} of the Spidron is the counterpart to the nonuniform folding in a periodic crease pattern; therefore, it would be a source of novel nonlinearity.
Another feature of the known Spidron's rigid folded state is the ``isotropy'', which means that the three diagonal distances in its hexagonal boundary are identical.
However, the hexagonal boundary of the Spidron itself has more DOF than $1$ as a linkage and can form a non-isotropic configuration.
Thus, questions arise; is the Spidron truly $1$-DOF, and is there no non-isotropic folded state?

In the present research, we focus on the non-self-similar folding of the recursive pattern which is the generalization of the Spidron with quadrilateral boundaries whose DOF is lower than that of the original hexagonal boundary, defined in Section~\ref{sec: crease-pattern}.
First, we describe the kinematics of a single unit cell of the Spidron in Section~\ref{sec: kinematics-single}.
To answer the question regarding the isotropy in our Spidron with quadrilateral boundaries, we do not assume that the two diagonal distances in a folded quadrilateral boundary are equal.
We will show that a single unit cell of the Spidron has $2$-DOF and can form a non-isotropic folded state.
Second, we analyze the kinematics of the Spidron with multiple unit cells in Section~\ref{sec: kinematics-multi}.
We will show that the non-isotropic folding is restricted as we increase the number of unit cells, which ensures that the Spidron behaves as almost $1$-DOF mechanism only allowing the isotropic folding.
Then, we analyze the isotropic folding using the dynamical system model and show their novel nonlinear nature, which includes instances of the classic period-doubling route to chaos.

\section{Crease Pattern of the Generalized Spidron}\label{sec: crease-pattern}
\begin{figure}[t]
    \centering
    \includegraphics[keepaspectratio,width=0.9\linewidth,page=3]{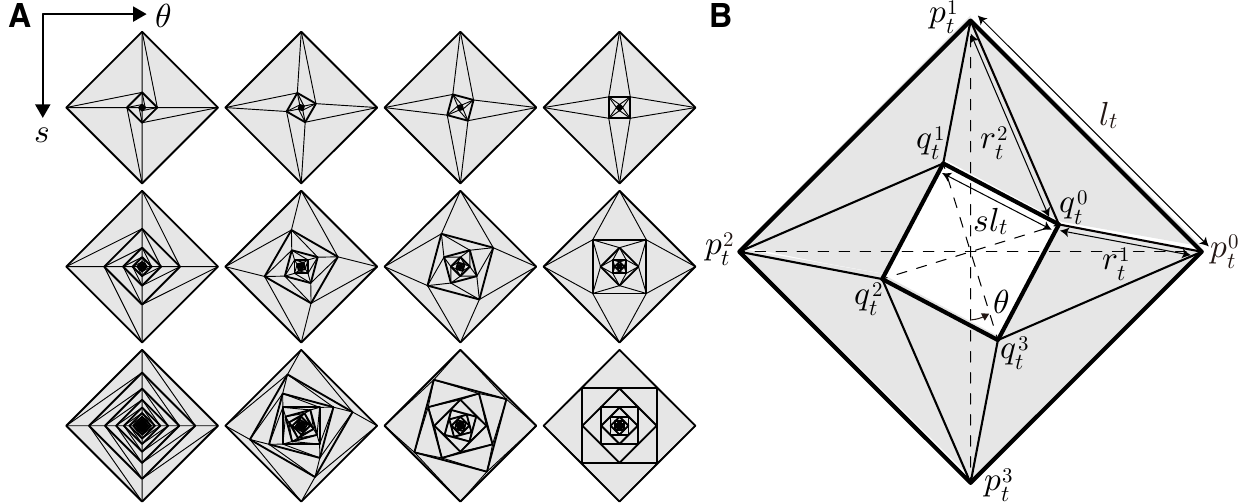}
    \caption{
        (A) Two-parameter family of the crease patterns for the generalized Spidron. 
        Each of the four rows, and columns shows $s=0.2, 0.5, 0.7$, and $\theta=0,\pi/12,\pi/6,\pi/4$, respectively.
        The crease patterns in the rightmost column are symmetric.
        (B) $t$-th ring of the generalized Spidron.
    }
    \label{fig: design}
\end{figure}

The crease pattern of the generalized Spidron is the $4$-fold symmetric recursive planar graph that can be obtained by repeatedly carrying out prescribed similarity transformation to an annular unit cell having square boundaries which we call a \emph{ring} (Figure~\ref{fig: design}).
The transformation is the scaling and rotating around the central point of squares, whose scaling factor and rotational angle are specified by two parameters $s\in(0,1)$ and $\theta\in[0,\pi/4]$, respectively.
Note that $\theta$ 
must be in the following range to avoid self-intersections of the creases:
\begin{align}
    0\leq\theta\leq\begin{cases}
        \frac{\pi}{4} &\text{if}\quad s\in\left(0,\frac{\sqrt{2}}{2}\right]\\
        \arccos{\left(\frac{1+\sqrt{2s^2-1}}{2s}\right)} &\text{if}\quad s\in\left[\frac{\sqrt{2}}{2},1\right)
    \end{cases}.
\end{align}
In addition, we do not assign mountain-valley labels to creases, at least explicitly.

In the following, we index rings by $t=0,1,\dots$ towards the inside, and write $p^{i}_t$, and $q^{i}_t= p^{i}_{t+1}$ for the vertices in their outer, and inner squares, respectively ($i\equiv 0,1,2,3 \pmod{4}$). 
Then, the lengths of creases $l_t:=|p^{i}_tp^{i+1}_t|$, $r^1_t:=|q^{i}_tp^{i}_t|$, and $r^2_{t}:=|q^{i}_tp^{i+1}_t|$ can be written down using $s$ and $\theta$ as follows:
\begin{align}
    l_t=s^t,\quad
    r^1_{t}=s^t\sqrt{\frac{1}{2}s^2-s\cos{\theta}+\frac{1}{2}},\quad
    r^2_{t}=s^t\sqrt{\frac{1}{2}s^2-s\sin{\theta}+\frac{1}{2}},\label{eq: edge-length}
\end{align}
where we fix the scale of $0$-th ring by $l_0=1$.
In the case of $\theta=\pi/4$ in which $r^1_t=r^2_t$ holds, the crease pattern has the additional mirror symmetries.
We call such a crease pattern \emph{symmetric}.

\section{Kinematics of a single ring}\label{sec: kinematics-single}
This section considers the kinematics of a single ring of the generalized Spidron.
The kinematics of rings are equivalent independent of their index $t$ because their crease patterns are scaled versions of each other.
Hence, we focus on the kinematics of the outermost $0$-th ring.
For the convenience of the notation, we omit the subscript $0$ for quantities such as edge lengths or vertices.
In addition, note that we allow self-intersections of facets to occur.
After parameterizing the folding of a single ring, we will consider its configuration space.
Then, we will consider folded states having a certain symmetry, which we call isotropic.
Note that we used Mathematica to derive and numerically solve the complex equations in this section.

\subsection{Parameterization}
\begin{figure}[t]
    \centering
    \includegraphics[keepaspectratio,width=0.9\linewidth]{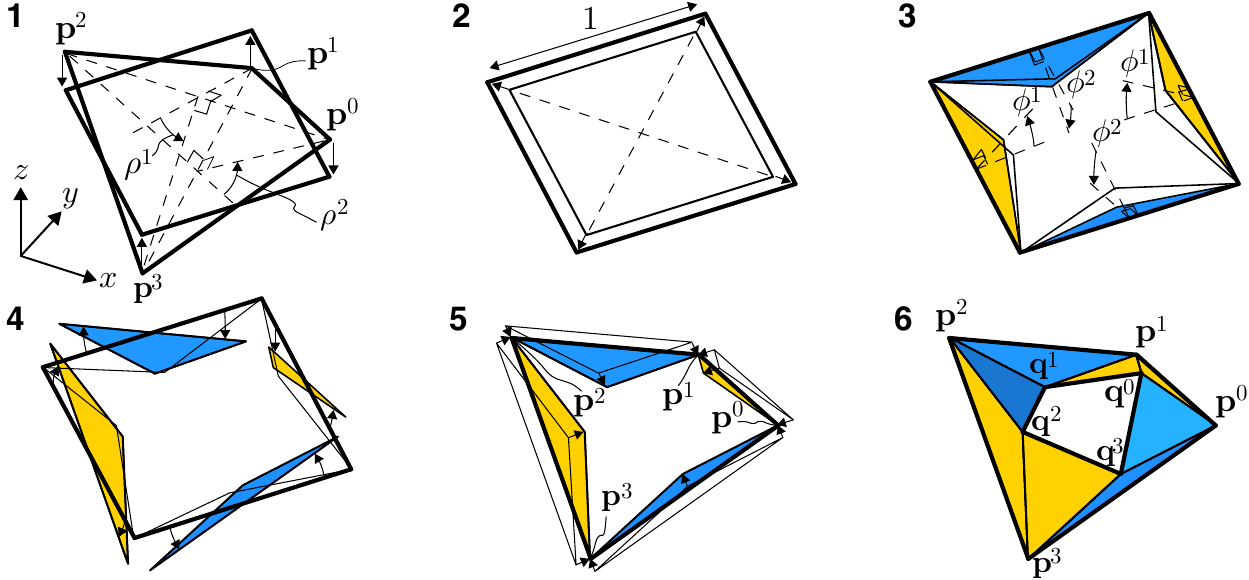}
    \caption{
        Construction of a folded state of a ring parameterized by $(\rho^1,\rho^2,\phi^1,\phi^2)$:
        (1) Project the outer boundary $\mathbf{p}^0\mathbf{p}^1\mathbf{p}^2\mathbf{p}^3$ specified by $\rho^1$ and $\rho^2$ to $xy$ plane.
        (2) Scale the projected boundary as it becomes the unit rhombus.
        (3) Rotate triangles lying on xy plane which has the metric prescribed by $\phi^1$ and $\phi^2$.
        (4) Rotate triangles around $\mathbf{v}^i$ so that their base edges become parallel with the edges of the outer boundary in (1).
        (5) Translate triangles so that their vertices match with $\mathbf{p}^i$.
        (6) If $|\mathbf{q}^0\mathbf{q}^1|=|\mathbf{q}^2\mathbf{q}^3|=s$ and $|\mathbf{q}^1\mathbf{q}^2|=|\mathbf{q}^3\mathbf{q}^0|=s$, $(\rho^1,\rho^2,\phi^1,\phi^2)$ represents a valid folded state of $0$-th ring.
    }
    \label{fig: parameterization}
\end{figure}
Let $\mathbf{p}^i$ and $\mathbf{q}^i\in\mathbb{R}^3$ denote the position vectors of vertices $p^i$ and $q^i$, respectively.
In order to represent the folded state of the ring, we will use four parameter angles $\rho^1, \rho^2 \in[-\pi,\pi]$, and $\phi^1, \phi^2 \in[-\pi,\pi]$ (Figure~\ref{fig: parameterization}).
The angles $\rho^1$ and $\rho^2$ are the exterior dihedral angles specifying the configuration of the outer boundary, which are formed by the oriented facets $\triangle \mathbf{p}^2\mathbf{p}^0\mathbf{p}^1$ and $\triangle \mathbf{p}^0\mathbf{p}^2\mathbf{p}^3$, $\triangle \mathbf{p}^1\mathbf{p}^3\mathbf{p}^0$ and $\triangle \mathbf{p}^3\mathbf{p}^1\mathbf{p}^2$, respectively.
We take the coordinate system so that $\mathbf{p}^i$ are parameterized as follows:
\begin{align}
    \mathbf{p}^0&=
    \begin{bmatrix}
        d^1\\
        0\\
        \frac{-\sgn{(\rho^1)}\sqrt{1-(d^1)^2-(d^2)^2}}{2}
    \end{bmatrix},\quad
    \mathbf{p}^1=
    \begin{bmatrix}
        0\\
        d^2\\
        \frac{-\sgn{(\rho^2)}\sqrt{1-(d^1)^2-(d^2)^2}}{2}
    \end{bmatrix},\nonumber\\
    \mathbf{p}^2&=\begin{bmatrix}
        -d^1\\
        0\\
        \frac{-\sgn{(\rho^1)}\sqrt{1-(d^1)^2-(d^2)^2}}{2}
    \end{bmatrix},\quad
    \mathbf{p}^3=\begin{bmatrix}
        0\\
        -d^2\\
        \frac{-\sgn{(\rho^2)}\sqrt{1-(d^1)^2-(d^2)^2}}{2}
    \end{bmatrix},\nonumber\\
    \text{where}&\quad d^i:=\frac{\left|\mathbf{p}^{i}\mathbf{p}^{i+2}\right|}{2}=\frac{\left|\tan{\frac{\rho^i}{2}}\right|}{\sqrt{\tan^2{\frac{\rho^1}{2}}+\tan^2{\frac{\rho^2}{2}}+\tan^2{\frac{\rho^1}{2}}\tan^2{\frac{\rho^2}{2}}}}\quad(i=1,2),
\end{align}
where $d^1,d^2\in[0,\sqrt{2}/2]$ are half the diagonal distances in the outer boundary.
Note that the signs of $\rho^1$ and $\rho^2$ must differ for $\left|\mathbf{p}^{i}\mathbf{p}^{i+1}\right|=1$ holds (In Figure~\ref{fig: parameterization} (1), $\rho^1<0$ and $\rho^2>0$.).
Next, we define $\mathbf{v}^i:=[0,0,1]^T\times(\mathbf{p}^{i+1}-\mathbf{p}^i)$.
The angle $\phi^1$ is the signed angle from $\mathbf{v}^0$ to $\triangle\mathbf{p}^0\mathbf{p}^1\mathbf{q}^0$ about the axis $\mathbf{p}^1-\mathbf{p}^0$, or from $\mathbf{v}^2$ to $\triangle\mathbf{p}^2\mathbf{p}^3\mathbf{q}^2$ about $\mathbf{p}^3-\mathbf{p}^2$.
In the same way, $\phi^2$ is the signed angle measured from $\mathbf{v}^1$ to $\triangle\mathbf{p}^1\mathbf{p}^2\mathbf{q}^1$ about $\mathbf{p}^2-\mathbf{p}^1$, or from $\mathbf{v}^3$ to $\triangle\mathbf{p}^3\mathbf{p}^0\mathbf{q}^3$ about $\mathbf{p}^0-\mathbf{p}^3$.

Then, the position vectors $\mathbf{q}^i$ can be written down using $\phi^1$ and $\phi^2$:
\small
\begin{align}
    \mathbf{q}^{0}&=
    \begin{bmatrix}
        \frac{
            d^1 \left(
                \sqrt{(d^1)^2+(d^2)^2}
                \left(1-(r^1)^2+(r^2)^2\right) \left((\tau^1)^2+1\right)
                +2 \sgn{(\rho^1)} S\sqrt{1-(d^1)^2-(d^2)^2}\tau^1
            \right)
            +d^2 S \left((\tau^1)^2-1\right)
        }
        {
            2 \sqrt{(d^1)^2+(d^2)^2}
            \left((\tau^1)^2+1\right)
        }\\
        \frac{
            d^2 \left(
                \sqrt{(d^1)^2+(d^2)^2}
                \left(1+(r^1)^2-(r^2)^2\right) \left((\tau^1)^2+1\right)
                -2 \sgn{(\rho^1)}S\sqrt{1-(d^1)^2-(d^2)^2}\tau^1
            \right)
            +d^1 S \left((\tau^1)^2-1\right)
        }
        {
            2 \sqrt{(d^1)^2+(d^2)^2}
            \left((\tau^1)^2+1\right)
        }\\
        \frac{1}{2}\sgn{(\rho^1)}\sqrt{1-(d^1)^2-(d^2)^2} \left((r^1)^2-(r^2)^2\right)
        +\frac{S\sqrt{(d^1)^2+(d^2)^2}\tau^1}{(\tau^1)^2+1}
    \end{bmatrix},\nonumber\\
    \mathbf{q}^{1}&=
    \begin{bmatrix}
        -\frac{
            d^1 \left(
                \sqrt{(d^1)^2+(d^2)^2}
                \left(1+(r^1)^2-(r^2)^2\right) \left((\tau^2)^2+1\right)
                +2 \mathrm{sgn}(\rho^1)S\sqrt{1-(d^1)^2-(d^2)^2}\tau^2
            \right)
            +d^2 S \left((\tau^2)^2-1\right)
        }
        {
            2 \sqrt{(d^1)^2+(d^2)^2}
            \left((\tau^2)^2+1\right)
        }\\
        \frac{
            d^2 \left(
                \sqrt{(d^1)^2+(d^2)^2}
                \left(1-(r^1)^2+(r^2)^2\right) \left((\tau^1)^2+1\right)
                -2 \mathrm{sgn}(\rho^1)S\sqrt{1-(d^1)^2-(d^2)^2}\tau^2
            \right)
            +d^1 S \left((\tau^2)^2-1\right)
        }
        {
            2 \sqrt{(d^1)^2+(d^2)^2}
            \left((\tau^2)^2+1\right)
        }\\
        \frac{1}{2}\sgn{(\rho^1)}
        \sqrt{1-(d^1)^2-(d^2)^2} 
        \left((r^2)^2-(r^1)^2\right)
        +\frac{S\sqrt{(d^1)^2+(d^2)^2}\tau^2}{(\tau^2)^2+1}
    \end{bmatrix},\nonumber\\
    \mathbf{q}^2&=[-q^0_x,-q^0_y,q^0_z]^T,\nonumber\\
    \mathbf{q}^3&=[-q^1_x,-q^1_y,q^1_z]^T.
\end{align}
\normalsize
where $\tau^i:=\tan{(\phi^i/2)}$, $S:=\sqrt{(r^1+r^2+1)(-r^1+r^2+1)(r^1-r^2+1)(r^1+r^2-1)}$ which is a quarter of the area of $\triangle p^ip^{i+1}q^i$.
For $(\rho^1,\rho^2,\phi^1,\phi^2)$ to represent a valid folded state with the same metric as the crease pattern, the following equations must hold:
\begin{align}
    \|\mathbf{q}^1-\mathbf{q}^0\|=
    \|\mathbf{q}^3-\mathbf{q}^2\|=s\quad \text{and} \quad 
    \|\mathbf{q}^2-\mathbf{q}^1\|=
    \|\mathbf{q}^0-\mathbf{q}^3\|=s.\label{eq: constraints}
\end{align}

\subsection{General Folding}
\begin{figure}[bt]
    \centering
    \includegraphics[keepaspectratio,width=0.85\linewidth]{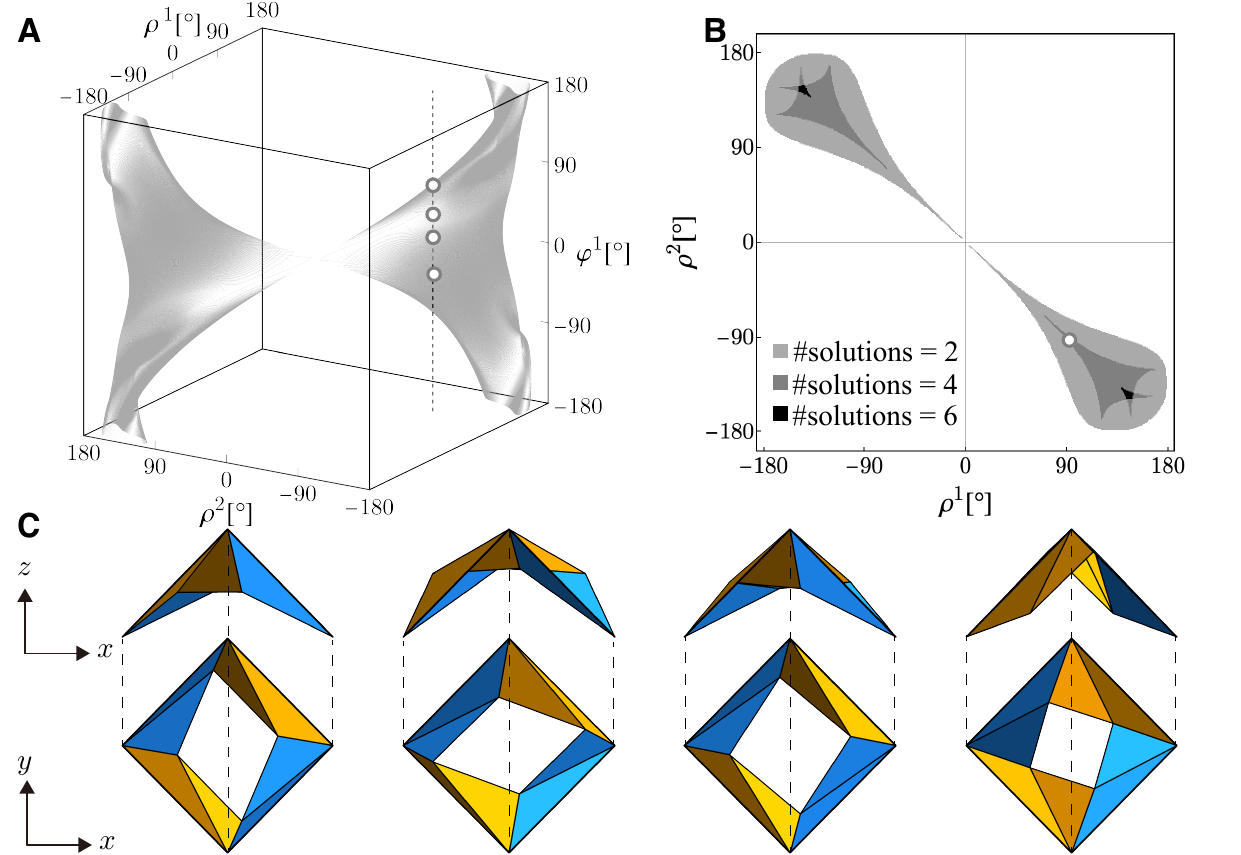}
    \caption{
        (A) Configuration space of a single ring in $(s,\theta)=(1/2,\pi/10)$ projected onto $(\rho^1,\rho^2,\phi^1)$ space. 
        (B) Projection of the configuration space to $(\rho^1,\rho^2)$ plane, where colors show the number of configurations compatible with the outer boundary corresponding to $(\rho^1,\rho^2)$.
        (C) Four folded states shown as circles in (A) and (B) whose outer boundaries are in the identical configuration.
    }
    \label{fig: configuration-space}
\end{figure}
The \emph{configuration space} of a single ring is the set of all its possible folded states and can be parameterized as the solution space of Equation~\eqref{eq: constraints} (Figure~\ref{fig: configuration-space} (A)).
As a simple calculation that the number of variables minus the number of constraints equals $2$ shows, the numerically calculated solution space of Equation~\eqref{eq: constraints} forms a $2$-dimensional surface in the parameter space; therefore, a single ring of the Spidron behaves as a $2$-DOF mechanism.
Note that the folded state of the ring is not uniquely determined only from the configuration of its outer boundary because Equation~\eqref{eq: constraints} can have multiple solutions of $\phi^1,\phi^2$ for some values of $\rho^1,\rho^2$ (Figure~\ref{fig: configuration-space} (B) and (C)).

In general, configuration spaces of origami have \emph{symmetry}, i.e., a transformation that takes a solution to another solution, representing the reversal of the front and back of a paper.
In our case, this symmetry can be represented as $(\overline{\rho}^1,\overline{\rho}^2,$$\overline{\phi}^1,$$\overline{\phi}^2)$ $\mapsto(-\overline{\rho}^1,-\overline{\rho}^2,$$-\overline{\phi}^1,$$-\overline{\phi}^2)$, where $(\overline{\rho}^1,\overline{\rho}^2,$$\overline{\phi}^1,$$\overline{\phi}^2)$ is a solution of Equation~\eqref{eq: constraints}.
Here, the $4$-fold symmetry of the crease pattern and the definition of the parameters give the solution space of Equation~\eqref{eq: constraints} an additional symmetry, which is represented as $(\overline{\rho}^1,\overline{\rho}^2,$$\overline{\phi}^1,$$\overline{\phi}^2)\mapsto(\overline{\rho}^2,\overline{\rho}^1,$$\overline{\phi}^2,$$\overline{\phi}^1)$.
Furthermore, if the crease pattern is symmetric as defined in Section \ref{sec: crease-pattern}, the solution space gets additional mirror symmetries: $(\overline{\rho}^1,\overline{\rho}^2,$$\overline{\phi}^1,$$\overline{\phi}^2)\mapsto(\overline{\rho}^2,\overline{\rho}^1,$$\overline{\phi}^1,$$\overline{\phi}^2)$ and $(\overline{\rho}^1,\overline{\rho}^2,$$\overline{\phi}^1,$$\overline{\phi}^2)\mapsto(\overline{\rho}^1,\overline{\rho}^2,$$\overline{\phi}^2,$$\overline{\phi}^1)$.

\subsection{Isotropic Folding}
\paragraph{Isotropic Folded State}
Here, we consider folded states whose parameter is invariant under the symmetry of the solution space of Equation~\eqref{eq: constraints} $(\rho^1,\rho^2,\phi^1,\phi^2)\mapsto(-\rho^2,-\rho^1,-\phi^2,-\phi^1)$, which is equivalent to $\rho^1=-\rho^2$ and $\phi^1=-\phi^2$.
We call such a folded state \emph{isotropic} because two diagonal distances in each of its outer and inner boundaries are the same.

\paragraph{Geometric Constraint in Isotropic Folding}
Geometric constraints in isotropic folding can be obtained by substituting $\rho^1=-\rho^2=\rho$ and $\phi^1=-\phi^2=\phi$ into Equation~\eqref{eq: constraints}.
Then, because of the $4$-fold symmetry of the crease pattern, the two equations  fall into a single equation, which is quartic in terms of $\tau:=\tan{(\phi/2)}$:
\small
\begin{align}
    &\left(
    \left(S^2-2 \left((r^1)^2+(r^2)^2-1\right)\right)d^2
    +\sqrt{2}S d
    +2 \left((r^1)^2+(r^2)^2\right)-s^2-S^2/2-1\right)\tau^4\nonumber\\
    +&2 \sgn{(\rho)}\sqrt{2} d \sqrt{1-2 d^2} \left((r^1)^2-(r^2)^2\right) S \tau ^3\nonumber\\
    +& \left(
    2\left(3 S^2-2 \left((r^1)^2+(r^2)^2-1\right)\right)d^2
    -S^2
    +2 \left(2 \left((r^1)^2+(r^2)^2\right)-s^2-1\right)\right)\tau ^2\nonumber\\
    +&2 \sgn{(\rho)}\sqrt{2} d \sqrt{1-2 d^2} \left((r^1)^2-(r^2)^2\right) S \tau\nonumber\\
    +&\left(
    \left(S^2-2 \left((r^1)^2+(r^2)^2-1\right)\right)d^2
    -\sqrt{2}S d
    +2 \left((r^1)^2+(r^2)^2\right)-s^2-S^2/2-1\right)=0,\label{eq: constraint-isotropic}
\end{align}
\normalsize
where $d:=d^1= d^2$ and $\tau:=\tau^1= -\tau^2$.

\begin{figure}[btp]
    \centering
    \includegraphics[keepaspectratio,width=\linewidth]{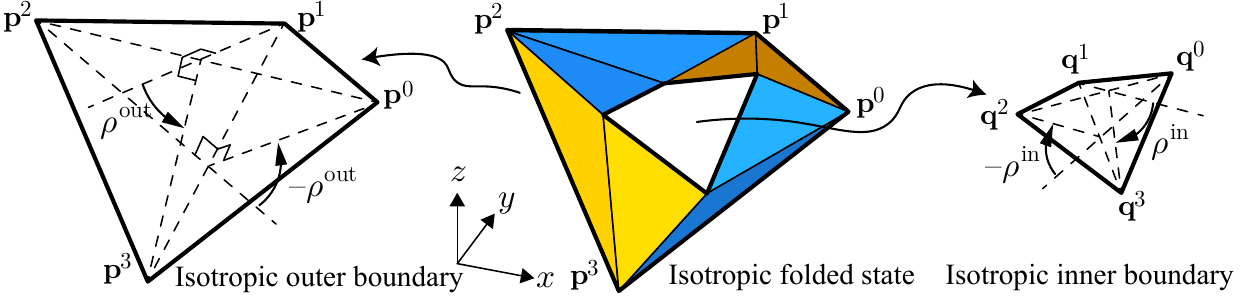}
    \caption{Reparameterization of an isotropic folded state by $(\rho^\mathrm{out}, \rho^\mathrm{in})$.}
    \label{fig: reparameterization}
\end{figure}

\begin{figure}[p]
    \centering
    \includegraphics[keepaspectratio,width=\linewidth]{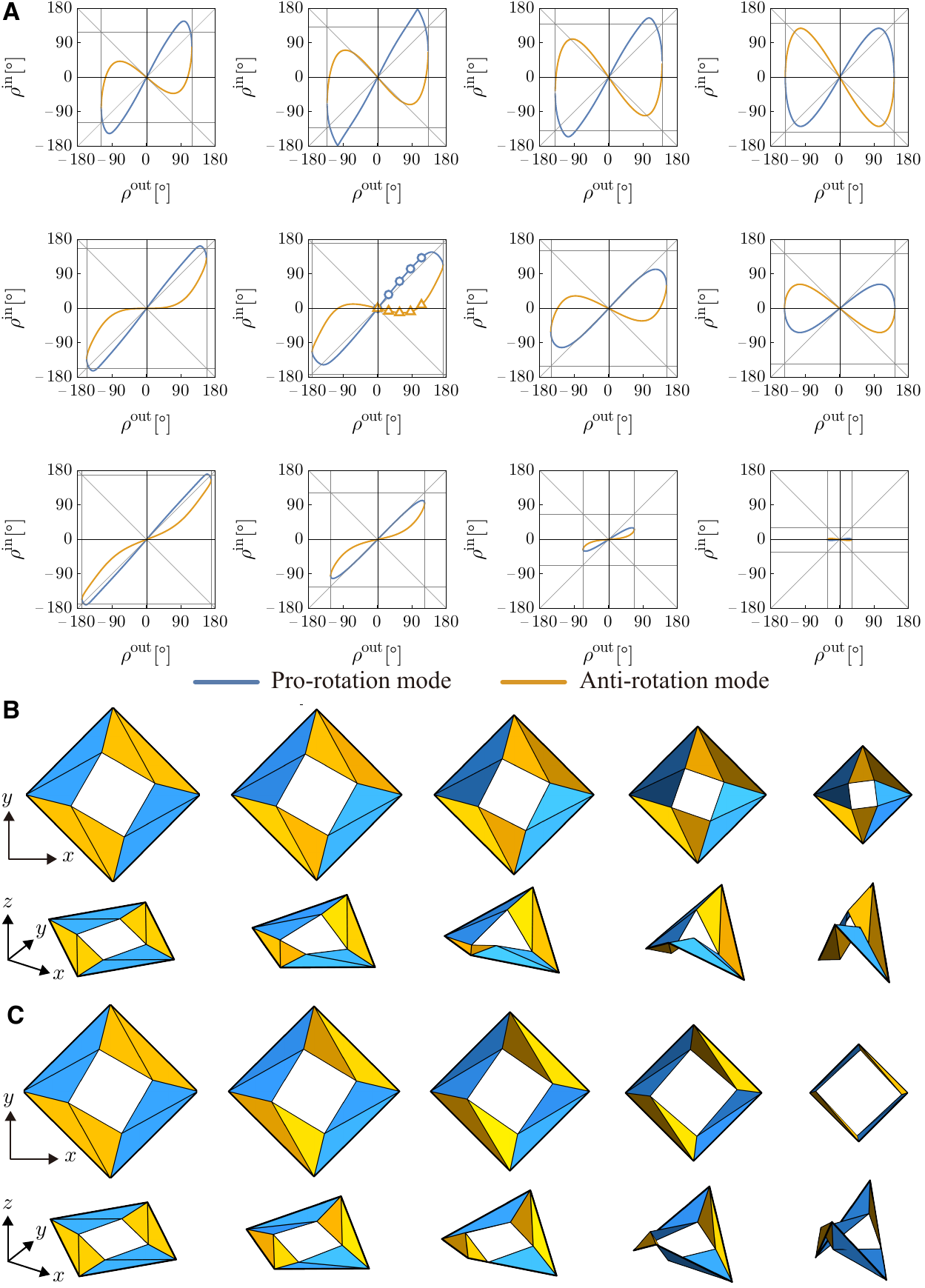}
    \caption{
        (A) Plots of $(\rho^{\mathrm{out}},\rho^{\mathrm{in}})$ representing isotropic folding motions of a ring for the crease patterns in Figure~\ref{fig: design} (A).
        The rightmost plots corresponding to the symmetric crease patterns are mirror symmetric about the horizontal axis.
        In each diagram, the boundary of $[\underline{x},\overline{x}]\times[\underline{x},\overline{x}]$ is indicated, where $[\underline{x},\overline{x}]$ is the domain of $\rho^{\mathrm{out}}$ in which $\rho^{\mathrm{in}}$ exists.
        (B)--(C) Pro-rotation, and anti-rotation mode in $(s,\theta)=(1/2,\pi/12)$.
        The points representing each of the folded states are plotted in (A).
        }
    \label{fig: plot_of_dynamical_systems}
\end{figure}

\paragraph{Two Isotropic Folding Modes}
Although a quartic equation can be solved using Ferrari's formula, the closed form of a solution is too long to be written down.
However, in the special case where a crease pattern is symmetric satisfying $r^1=r^2$, the first and third-order terms vanish, and Equation~\eqref{eq: constraint-isotropic} becomes simpler:
\begin{multline}
    \left(
    d^2
    -(2s-\sqrt{2})d
    -\sqrt{2}s+1/2\right)\tau^4
    +\left(
    2\left(4s^2-4\sqrt{2}s+3\right)d^2
    -2\sqrt{2}s+1
    \right)\tau ^2\\
    +\left(
    d^2
    +(2s-\sqrt{2})d
    -\sqrt{2}s+1/2\right)=0,\label{eq: constraint-isotropic-symmetric}
\end{multline}
where we assumed $s\in(0,\sqrt{2}/2)$.
Then, we can write down the solution of the equaiton~\eqref{eq: constraint-isotropic-symmetric} in the following relatively simple form:
\small
\begin{multline}
    \tau=
    \pm\left(\left(
        \pm4 \sqrt{d^2 \left(4 d^2 s^4-2 \sqrt{2} \left(4 d^2+1\right) s^3+2 \left(7 d^2+3\right) s^2-3 \sqrt{2} \left(2 d^2+1\right) s+2 d^2+1\right)}\right.\right.\\
        \left.\left.
        -\left(
        2\left(4s^2-4\sqrt{2}s+3\right)d^2
        -2\sqrt{2}s+1
        \right)
    \right)/
    \left(2d^2-2 d \left(2 s-\sqrt{2}\right)-2 \sqrt{2} s+1\right)
    \right)^{1/2}.\label{eq: solution_of_constraints_isotropic_symmetric}
\end{multline}
\normalsize
Here, at most two of the four solutions~\eqref{eq: solution_of_constraints_isotropic_symmetric} can be real.
Therefore, a symmetric crease pattern has two $1$-parameter families of isotropic folded states, i.e., \emph{isotropic folding motions} which branch from the developed state.
Let us reparameterize the isotropic folded state by two signed dihedral angles $\rho^\mathrm{out}$ and $\rho^\mathrm{in}$ which are measured in the same way as $\rho$, representing the isotropic configuration of the outer and inner boundary, respectively (Figure~\ref{fig: reparameterization}).
Then, if we write $\tau(\rho)$ for the real solution of the equation for the given $\rho$, $\rho^{\mathrm{in}}$ is given as the function of $\rho^{\mathrm{out}}=\rho$ as follows:
\begin{align}
    \tan{\frac{\rho^{\mathrm{in}}}{2}}=\frac{2q^{1}_z(\rho^{\mathrm{out}},\tau(\rho^{\mathrm{out}}))}{\sqrt{q^{1}_{x}(\rho^{\mathrm{out}},\tau(\rho^{\mathrm{out}}))^2+q^{1}_{y}(\rho^{\mathrm{out}},\tau(\rho^{\mathrm{out}}))^2}}.\label{eq: outer-inner-boundary}
\end{align}
We observed the existence of two isotropic folding modes also in the non-symmetric crease patterns (Figure~\ref{fig: plot_of_dynamical_systems} (A)).
These modes may be characterized using the notion of pop-up/down assignments on interior vertices~\cite{chen2018branches}.
Intuitively, we can distinguish the two modes by the rotation direction of the inner boundary relative to the outer boundary viewed from the top (Figure~\ref{fig: plot_of_dynamical_systems} (B) and (C)).
If the inner boundary is rotated anti-clockwise/clockwise, we call these modes \emph{pro-}/\emph{anti-rotation mode}, based on the fact that the inner boundary is rotated anti-clockwise relative to the outer boundary on non-symmetric crease patterns.
In the following, we write $f_\mathrm{pro}$ and $f_\mathrm{anti}$ for the mappings which map $\rho^\mathrm{out}$ to $\rho^\mathrm{in}$ representing the pro- and anti-rotation mode, respectively.
In addition, let $[\underline{x},\overline{x}]$ denote the domain of these mappings.

\section{Kinematics of Multiple Rings}\label{sec: kinematics-multi}
This section considers the kinematics of the Spidron with multiple rings based on the following recurrence construction.
Given a folded state of the $t$-th ring, its inner boundary serves as the outer boundary of the $(t+1)$-st ring.
Then, we can obtain a folded state of the $(t+1)$-st ring that is compatible with its outer boundary by solving the geometric constraints represented by Equation~\eqref{eq: constraints} or \eqref{eq: constraint-isotropic}.
The entire folded state can be constructed by carrying out the above procedure iteratively.
 Same as Section~\ref{sec: kinematics-single}, we used Mathematica to numerically solve the equations.

Firstly, we will show that the Spidron with multiple rings behaves as an ``almost'' $1$-DOF mechanism that only exhibits the isotropic folding motions.
Although a single ring behaves as a $2$-DOF mechanism, as the number of rings increases, non-isotropic folding is restricted because of the geometric constraints that come from each of the rings.
Hence, we 
focus on the isotropic folding.
We will analyze it using a dynamical system representing the recurrence construction, and show its nonlinear nature including the period-doubling bifurcation.

\subsection{General Folding: Convergence to Isotropic Folding}

\paragraph{Parameterization}
Let $\rho^{i}_t,\phi^{i}_t$ denote the parameters representing the folded state of $t$-th ring, whose definition was given in Section~\ref{sec: kinematics-single}.
Then, the folded state of the Spidron with $T\in\mathbb{Z}_{>0}$ rings can be parameterized by $\{(\rho^1_t,\rho^2_t,\phi^1_t,\phi^2_t)\}_{t=0,1,\dots,T-1}$ where Equation~\eqref{eq: constraints} holds for every $t$, and adjacent terms are related by the following equation:
\begin{align}
    \tan\frac{\rho^{1}_{t+1}}{2}=
    \frac
    {q^{1}_{t,z}-q^{0}_{t,z}}
    {\left|\mathbf{q}^{1}_{t}-\mathbf{q}^{3}_{t}\right|/2},\quad
    \tan\frac{\rho^{2}_{t+1}}{2}=
    \frac
    {q^{0}_{t,z}-q^{1}_{t,z}}
    {\left|\mathbf{q}^{2}_{t}-\mathbf{q}^{0}_{t}\right|/2}.
\end{align}

\begin{figure}[t]
    \centering
    \includegraphics[keepaspectratio,width=\linewidth]{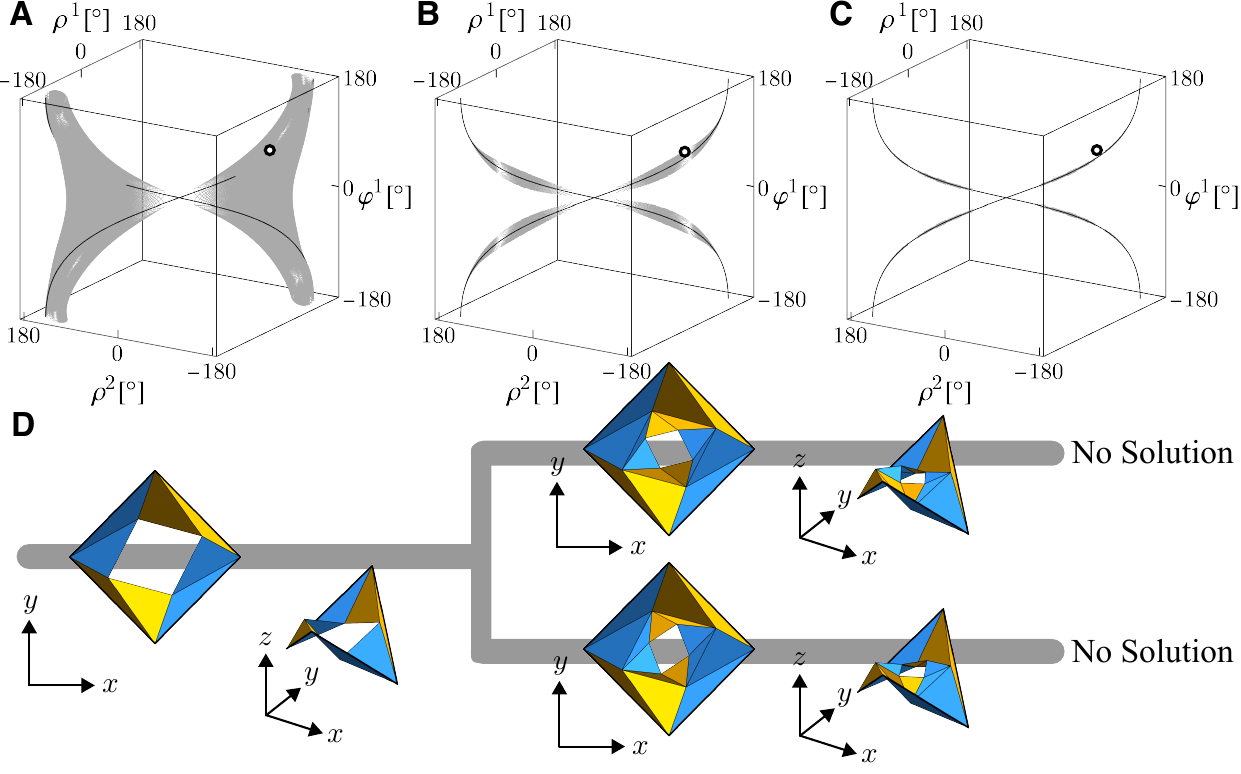}
    \caption{
        Configuration space of multiple rings and recursive construction of a folded state under the fixed crease pattern $(s,\theta)=(0.45,\pi/4)$.
        (A)--(C) Configuration space of single, double, and triple rings, respectively.
        The black curve on the gray surface shows the isotropic folding modes of a single ring.
        (D) Recursive construction from a folded state represented by $(\rho^1_0,\rho^2_0,\phi^1_0,\phi^2_0)=(1.847, -1.860, 1.140, -0.768)$ plotted as black circles in (A)--(C).
        For this $0$-th folded state, there are two different folded states for the $1$-st ring.
        Then, there is no folded state of the $2$-nd ring compatible with either of the $1$-st ones.
    }
    \label{fig: configuration-space_multi}
\end{figure}

\paragraph{Configuration Space of Multiple Rings}
Once we set values of $\rho^1_0$ and $\rho^2_0$, we can get $\{(\rho^1_t,$ $\rho^2_t,$ $\phi^1_t,$ $\phi^2_t)\}_{t=0, \dots, T-1}$ representing a folded state of the Spidron with $T$ rings through solving Equation~\eqref{eq: constraints} iteratively.
Then, we can project the configuration space of the Spidron with $T$ rings to that of the $0$-th ring by projecting a sequence $\{(\rho^1_t,\rho^2_t,\phi^1_t,\phi^2_t)\}_{t=0, \dots, T-1}$ to $(\rho^1_0,\rho^2_0,\phi^1_0)$ space (Figure~\ref{fig: configuration-space_multi}).
Note that some sequences can be projected onto the identical point because there can be multiple configurations compatible with the given outer boundary as we saw in Section~\ref{sec: kinematics-single}.

\paragraph{Degeneracy of Degrees of Freedom}
Through the numerical calculation, we observed that as the number of rings increases, the projected configuration space shrinks to the curve corresponding to the isotropic folding modes (Figure~\ref{fig: configuration-space_multi} (A)--(C)).
One typical behavior in the recursive construction is that given a folded state of the initial ring whose outer boundary is in a non-isotropic configuration, the degree of non-isotropy of boundaries increases as we go inwards, and finally, a recursive construction fails (Figure~\ref{fig: configuration-space_multi} (D)).
This result shows the degeneracy of degrees of freedom from $2$ to $1$, which means the restriction of the non-isotropic folding and the dominance of the isotropic folding in the Spidron with multiple rings.
Thus, we will focus on the isotropic folding in the subsequent part.

\subsection{Isotropic Folding}
\paragraph{Parameterization}
We will call a folded state of the Spidron with $T$ rings \emph{isotropic} if all the $T$ rings are in an isotropic folded state.
The isotropic  folded state can be parameterized by $\{(\rho^{\mathrm{out}}_t,\rho^{\mathrm{in}}_t)\}_{t=0,1,\dots,T-1}$ where Equation~\eqref{eq: outer-inner-boundary} holds for each $t$, and adjacent terms are related by $\rho^{\mathrm{out}}_{t+1}=\rho^{\mathrm{in}}_{t}$.

\begin{figure}[p]
    \centering
    \includegraphics[keepaspectratio,width=\linewidth,page=2]{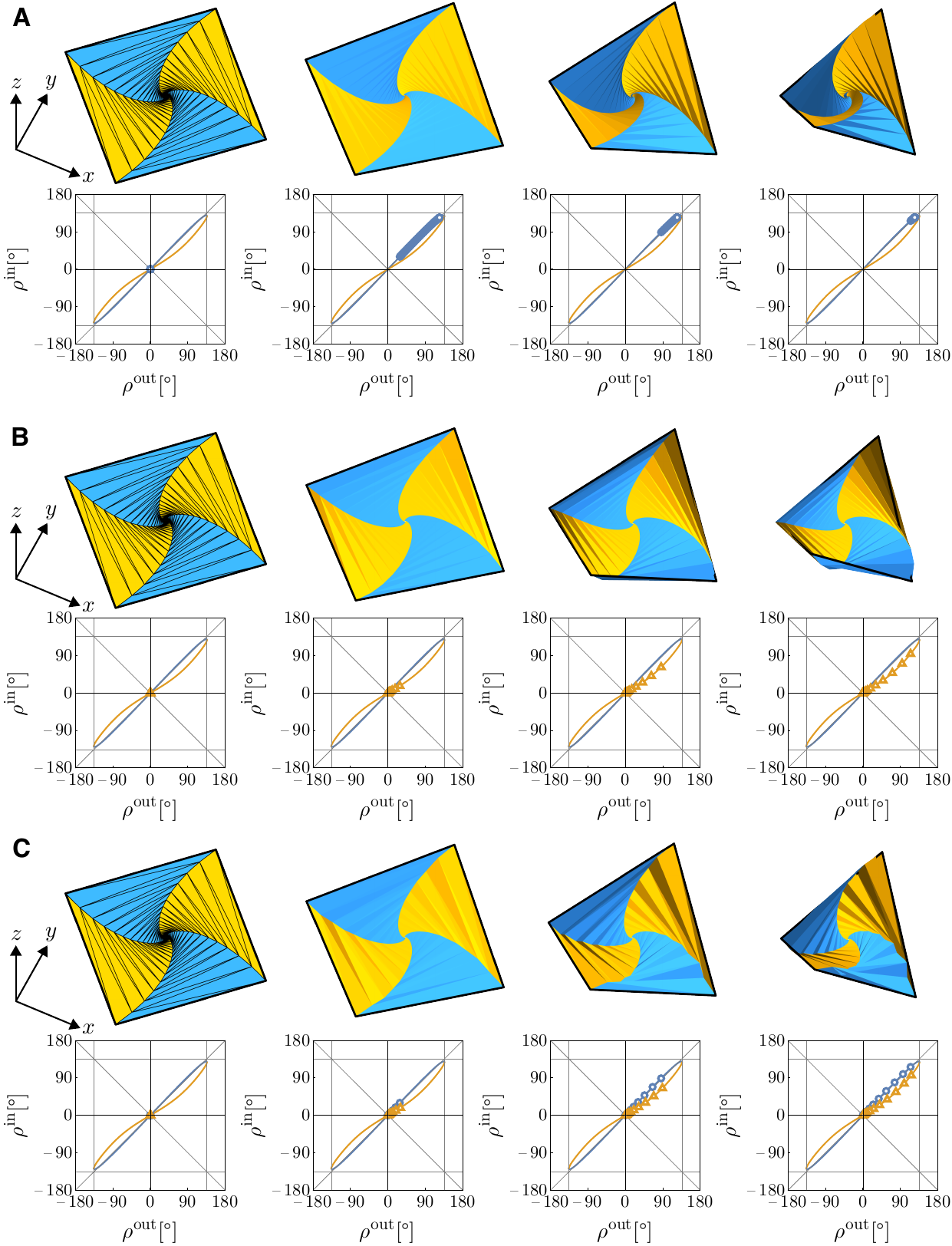}
    \caption{
        Isotropic folding motions with the three different mode assignments under the fixed crease pattern $(s,\theta)=(0.825,0.09)$: (A) pro-rotation mode, (B) anti-rotation mode, (C) pleats mode.
        Each figure in (A)--(C) shows the folded state for $\rho^{\mathrm{out}}_0=0,0.5,1.5$, and $2.0$ from left to right, respectively.
        In each figure, the solution $\{(\rho^{\mathrm{out}}_t,\rho^{\mathrm{in}}_t)\}_{T=0,\dots,30}$ corresponding to the folded state is plotted on the graph of $(\rho^\mathrm{in},\rho^\mathrm{out})$ for $(s,\theta)=(0.825,0.09)$.
    }
    \label{fig: three-modes}
\end{figure}

\begin{figure}[t]
    \centering
    \includegraphics[keepaspectratio,width=\linewidth,page=3]{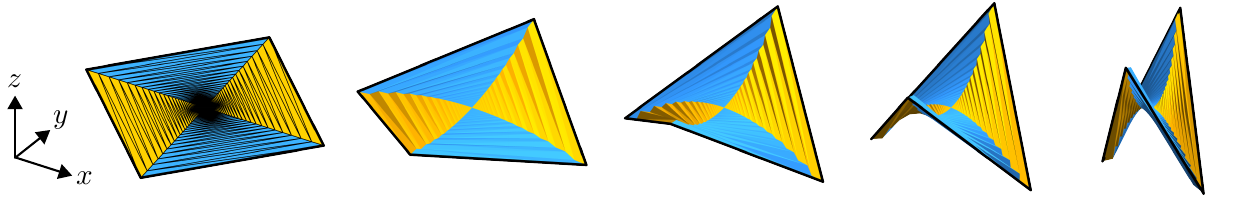}
    \caption{
        Pleats mode in the crease pattern $(s,\theta)=(0.9,0)$ which is equivalent to the symmetrically triangulated hypar origami whose central square is removed.
    }
    \label{fig: hypar}
\end{figure}

\paragraph{Mode Assignment}
As we saw in Section~\ref{sec: kinematics-single}, for the given $\rho^\mathrm{out}_t\in[\overline{x},\underline{x}]$, there are two different candidates for $\rho^\mathrm{in}_t$ which are represented by $f_\mathrm{pro}(\rho^\mathrm{out}_t)$ and $f_\mathrm{anti}(\rho^\mathrm{out}_t)$; therefore, there is a degree of freedom in \emph{mode assignment} for each ring.
Hence, assuming the existence of solutions at every $t$, the number of possible folding modes of the entire structure increases exponentially as we increase the number of rings.
Here, we consider prescribing mode assignments in the following three ways: 1) \emph{pro-rotation mode}, where all rings must take the pro-rotation mode, which corresponds to the known isotropic folding motion of the original Spidron~\cite{szilassi2004right,kiss2018spidron}, 2) \emph{anti-rotation mode}, where all rings must take the anti-rotation mode, and 3) \emph{pleats mode}, where the rings take the pro- and anti-rotation mode alternately.
In particular, we consider the case in which $2t$-th and $(2t+1)$-st ring must take the pro-rotation and anti-rotation mode, respectively (Figure~\ref{fig: three-modes}).
The pleats mode is similar to the $1$-DOF rigid folding motion of the well-known triangulated hypar origami~\cite{demaine2011non,filipov2018mechanical,liu2019invariant}.
The most similar case is when $\theta=0$, where some creases become colinear as in the hypar pattern (Figure~\ref{fig: hypar}).
However, our crease pattern or its folding motion differs slightly from that even if $\theta=0$.
This is because the triangulation of the crease pattern in the previous works does not have 4-fold symmetry, and there is a diagonal crease on its central square, which makes its rigid folding motion non-isotropic.

\paragraph{Dynamical System}
Recurrence behaviors under the prescribed mode assignment can be qualitatively/quantitatively understood using a dynamical system model.
Owing to the mode-assignment for each ring, we can uniquely determine $\rho^{\mathrm{in}}_t=\rho^{\mathrm{out}}_{t+1}$ from $\rho^{\mathrm{out}}_t$.
In particular for the pro- and anti-rotation modes, the mapping $\rho^{\mathrm{out}}_t\mapsto \rho^\mathrm{out}_{t+1}$ is given by $f_\mathrm{pro}$ and $f_\mathrm{anti}$ for every $t$.
Thus, $f_\mathrm{pro}$ and $f_\mathrm{anti}$ serve as the recurrence relation in the pro- and anti-rotation mode, respectively, which are \emph{$1$-dimensional discrete dynamical systems}.
For the pleats mode, we can define its dynamical system $f_{\mathrm{pleats}}$ as $f_{\mathrm{anti}}\circ f_{\mathrm{pro}}:\rho^{\mathrm{out}}_t\mapsto\rho^\mathrm{out}_{t+2}$.
Note that if $f_{\sigma}([\underline{x},\overline{x}])\subseteq[\underline{x},\overline{x}]$ and $\rho_0^{\mathrm{out}}\in[\underline{x},\overline{x}]$, $\rho^{\mathrm{out}}_t$ exists for all $t$, which ensures the existence of a corresponding folded state ($\sigma\in\{\mathrm{pro},\mathrm{anti},\mathrm{pleats}\}$).
This condition can be judged visually by checking whether the graph of $f_\sigma$ is included in a square representing $[\overline{x},\underline{x}]\times[\overline{x},\underline{x}]$ or not (Figure~\ref{fig: plot_of_dynamical_systems}).

\paragraph{Recurrence Behavior}
Even if the crease patterns are identical, recurrence behaviors can change dramatically depending on mode assignments because different modes are governed by different dynamical systems.
If we increase the number of rings infinitely, the behavior of an \emph{orbit} $\{\rho^\mathrm{out}_t\}_{t=0,1,\dots}$ in the limit $t\rightarrow\infty$ is significant.
One typical behavior is the convergence to a \emph{stable fixed point} satisfying $f_\sigma(\rho^\mathrm{out})=\rho^\mathrm{out}$ and $|f'_{\sigma}(\rho^\mathrm{out})|<1$.
Note that the folded state corresponding to an orbit initiated from a fixed point exhibits a self-similarity whose scaling ratio is the same as the crease pattern.
This self-similarity is because the orbit of a fixed point forms a constant sequence, meaning that all the rings in the folded state are similar.
Graphically, we can detect fixed points by taking the intersection of the graphs of $\rho^\mathrm{in}=\rho^\mathrm{out}$ and  $f_\sigma(\rho^\mathrm{out})$ (Figure~\ref{fig: plot_of_dynamical_systems}).
For example, $\rho^\mathrm{out}=0$ corresponding to the developed state is always a fixed point of $f_{\sigma}$ independent of crease pattern and $\sigma$, and the orbits in Figure~\ref{fig: three-modes} (B) and (C) converge to this point.
Such convergence to a developed state has been reported in previous work~\cite{szilassi2004right}.
However, in Figure~\ref{fig: three-modes} (A), the orbits converge to the non-trivial fixed point corresponding to the finitely folded state rather than the flat state that is unstable there.
Such non-trivial behavior has not been reported in previous works, and it becomes possible because we generalized the crease pattern.

\paragraph{Bifurcation Analysis}
\begin{figure}[t]
    \centering
    \includegraphics[keepaspectratio,width=0.9\linewidth]{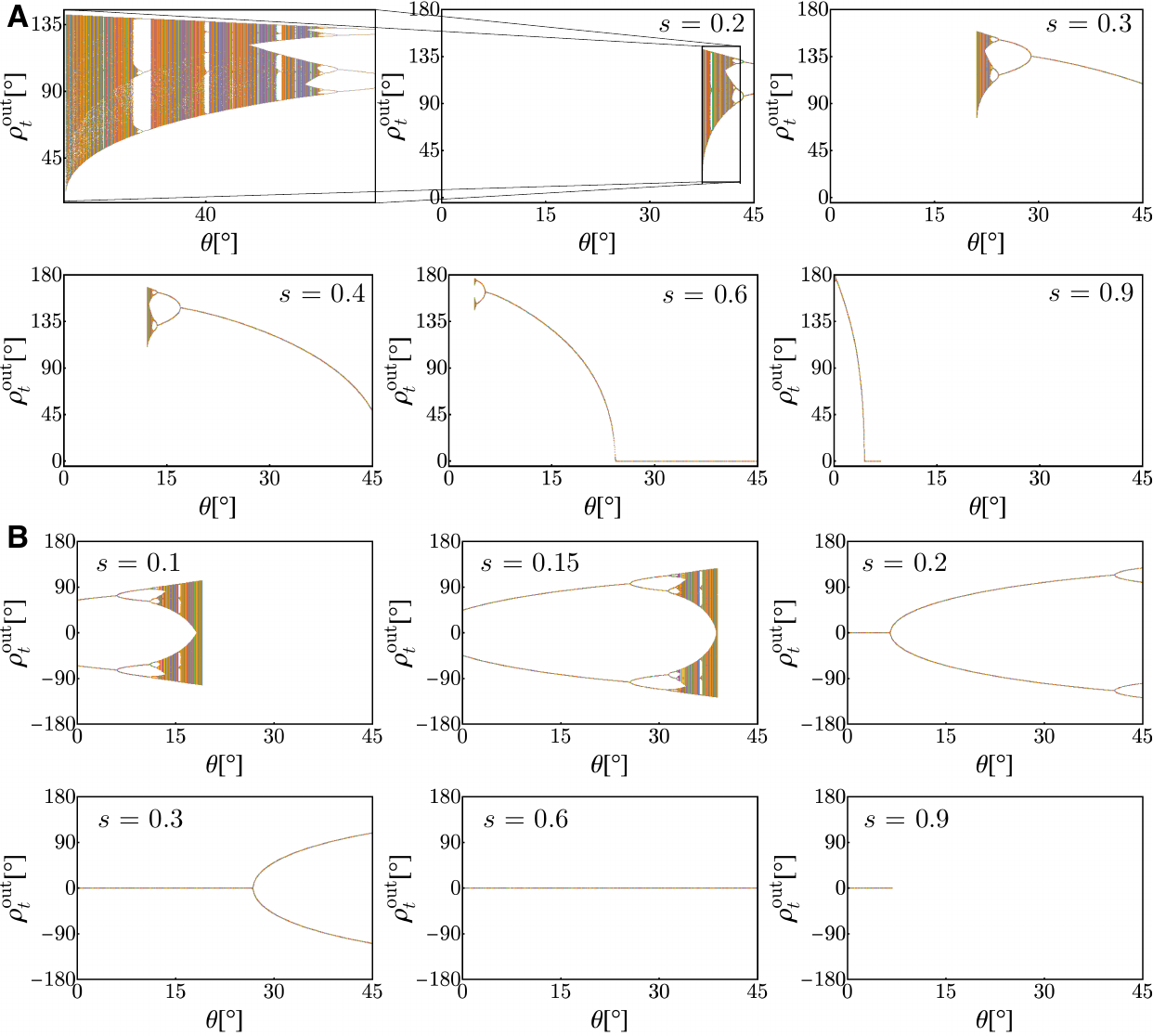}
    \caption{
        (A), (B) Bifurcation diagrams for $f_\mathrm{pro}$ and $f_\mathrm{anti}$ under the fixed values of $s$.
        For each value of $\theta$ satisfying the sufficient condition for infinite repetition, we randomly chose $\rho^\mathrm{out}_0\in[\underline{x},\overline{x}]$ and plotted $\{(\theta, \rho^\mathrm{out}_t)\}_{t=4001,\dots,5000}$.
    }
    \label{fig: bifurcation}
\end{figure}
Tuning of the crease pattern affects the recurrence behavior of the dynamical systems.
We observed the change of recursive behaviors of $f_\mathrm{pro}$ and $f_\mathrm{anti}$ by carrying out the bifurcation analysis where we fix the value of $s$ and change the value of $\theta$ (Figure~\ref{fig: bifurcation}).
Some of the bifurcation diagrams show \emph{period-doubling behavior} which leads to the emergence of \emph{chaos}.
Basically, both $f_\mathrm{pro}$ and $f_\mathrm{anti}$ exhibit non-trivial recursive behavior in small $s$, where ``non-trivial'' means the convergence to a fixed point corresponding to a finitely folded state, periodic points, or a chaotic attractor.
For fixed $s$, $f_\mathrm{pro}$ exhibits non-trivial behavior in relatively small $\theta$ while $f_\mathrm{anti}$ does so in relatively large $\theta$.
Note that self-intersections of faces are likely to occur when $\rho^{\mathrm{out}}_t$ is large; therefore, we have to find a chaotic solution whose upper/lower value is small to avoid the self-intersections.
Such a highly nonlinear structure of origami has not been reported in previous works and will open the way for new possibilities of origami structures.

\section{Conclusion}\label{sec:conclusion}
In this paper, we presented the nonlinear kinematics of the generalized Spidron.
In particular, we observed the degeneracy of DOF, the convergence of orbits to stable points in the dynamical systems dominating the isotropic folding, and the period-doubling route to chaos.
These phenomena differ from the periodic type of origami patterns without scaling, which is primarily governed by quasi-periodic undulations~\cite{imada2023undulations}.
They are caused by the system's asymmetry with respect to the timestep directions due to scaling; i.e., whether taking the next step inward or outward changes the recurrence relation. This asymmetry also results in the non-conservativeness of the dynamical system.
The method and results presented here will be useful in understanding and analyzing various recursive origami patterns, not limited to the Spidron.

The isotropic rigid folding of the Spidron that converges to the finitely folded state could be used to store sheets in the compact configuration as the well-known flasher origami, which is not rigidly foldable without cuts or the triangulation~\cite{lang2016single}.
The quick convergence to a finitely folded state implies that the structure folds to a desired shape automatically by just actuating the outermost boundary of the developed structure.
In addition, we observed that a real model of the Spidron can have multi-stability as the hypar origami~\cite{filipov2018mechanical,liu2019invariant}, which is caused by the mode-switching between pro- and anti-rotation modes.
Because mode-switching results in the switching of the dominating dynamical system, it can switch the mechanical behavior of the structure.
From these engineering perspectives, exploring the mechanical properties of our pattern is one of the possible future directions.
Another direction would be the generalization of the boundary shapes to such as rectangular~\cite{demaine1999polyhedral}, hexagonal boundaries~\cite{chen2023multi} like the original Spidron, or circular boundaries.
This generalization would bring new mathematical/mechanical properties.

There are some more problems regarding recursive patterns that would be interesting if we could solve them using the recursive nature.
For example, when creating and folding a real model, considering the self-foldability~\cite{tachi2017self,chen2018branches}, meaning taming the bifurcation at the unfolded state, would be inevitable.
For another direction, using recursive patterns in the boundary-filling problem not only for a prescribed folded state of a boundary~\cite{demaine2015filling} but for a prescribed folding motion~\cite{lee2024designing} would also be interesting.

\section*{Acknowledgement} 
The question leading to this work was first proposed and discussed in Structural Origami Gathering 2017.
This work is supported by JSPS KAKENHI Grant Number JP23KJ0682.
Author TCH is supported by NSF grant DMS-2347000.

\bibliographystyle{osmebibstyle}
\bibliography{osmerefs}

\theaffiliations

\end{document}